\documentclass[epj,twocolumn]{webofc} 
\usepackage[varg]{txfonts}   % Web of Conferences font
\usepackage{graphicx}% Include figure files
\newcommand{\ba}{\begin{eqnarray}}
\newcommand{\ea}{\end{eqnarray}}
\newcommand{\bsub}{\begin{subequations}}
\newcommand{\esub}{\end{subequations}}
\woctitle{CGS15}
\begin{document}
\selectlanguage{english}
\title{Partial dynamical symmetry and odd-even staggering 
in deformed nuclei}
%
% subtitle is optionnal
%
%%%\subtitle{Do you have a subtitle?\\ If so, write it here}

\author{A. Leviatan\inst{1}\fnsep\thanks{\email{ami@phys.huji.ac.il}}}

\institute{Racah Institute of Physics, The Hebrew University, 
Jerusalem 91904, Israel}

\abstract{
Partial dynamical symmetry (PDS) is shown to be relevant for describing 
the odd-even staggering in the $\gamma$-band of $^{156}$Gd while retaining 
solvability and good SU(3) symmetry for the ground and $\beta$ bands.
Several classes of interacting boson model Hamiltonians with SU(3) PDS 
are surveyed.}
\maketitle
A convenient starting point for describing axially-deformed nuclei 
is the SU(3) limit of the interacting boson model 
(IBM)~\cite{Iachello87}. 
The latter limit corresponds to the chain of nested algebras 
\ba
\begin{array}{cccccc}
{\rm U}(6)&\supset&{\rm SU}(3)&
\supset&{\rm SO}(3)\\
\downarrow&&\downarrow&&\downarrow&\\[0mm]
[N]&&(\lambda,\mu)&K&L
\end{array},
\label{e_clas}
\ea
where below each algebra the associated labels of irreducible 
representations (irreps) are given, and $K$ is a multiplicity label. 
The eigenstates $\vert [N](\lambda,\mu)K,L\rangle$
are obtained with a Hamiltonian
with SU(3) DS which has the form
\begin{equation}
\hat H_{\rm DS}=
\alpha_1\hat C_2[{\rm SU(3)}]+
\alpha_2\hat C_2[{\rm SO(3)}]+
\alpha_3\hat C_3[{\rm SU(3)}] ~.
\label{e_ds}
\end{equation}
The quadratic and cubic Casimir operators of SU(3) are
$\hat C_2[{\rm SU(3)}] = 
2\hat{Q}\cdot \hat{Q} + {\textstyle\frac{3}{4}}\hat{L}\cdot \hat{L}$ 
and $\hat C_3[{\rm SU(3)}] =
-4\,\sqrt{7}\hat{Q}\cdot (\hat{Q}\times \hat{Q}) ^{(2)}
-{\textstyle\frac{9}{2}\sqrt{3}}\hat{Q}\cdot 
(\hat{L}\times \hat{L})^{(2)}$, and  
$\hat C_2[{\rm SO(3)}]=\hat{L}\cdot \hat{L}$.
The SU(3) generators are $\hat{Q} = d^{\dagger}s + s^{\dagger}\tilde{d} 
-{\textstyle\frac{1}{2}}\sqrt{7}\, (d^{\dagger}\tilde{d})^{(2)}$ and 
$\hat{L} = \sqrt{10}\, (d^{\dagger}\tilde{d})^{(1)}$.
The monopole $(s)$ 
and quadrupole $(d)$ bosons represent valence nucleon pairs 
whose total number $N$ is conserved. 
$\hat{H}_{DS}$ is completely solvable with eigenenergies
\ba
E_{\rm DS} = \alpha_{1}f_{2}(\lambda,\mu) + \alpha_{2}L(L+1) 
+\alpha_{3}f_{3}(\lambda,\mu) ~,
\label{ene_ds}
\ea
where $f_{2}(\lambda,\mu)\!=\!\lambda^2\!+\!(\lambda\!+\!\mu)(\mu\!+\!3)$ 
and $f_{3}(\lambda,\mu)\!=\!(\lambda\!-\!\mu)(2\lambda\!+\!\mu\!+\!3)
(\lambda\!+\!2\mu\!+\!3)$. 
The spectrum~resembles 
that of a quadrupole axially-deformed rotor with eigenstates 
arranged in SU(3) multiplets and $K$ corresponds geometrically 
to the projection of the angular momentum on the symmetry axis. 
The lowest SU(3) irrep $(2N,0)$ contains the ground band g$(K\!=\!0)$,  
while $(2N-4,2)$ encompasses the $\beta(K\!=\!0)$ and 
$\gamma(K\!=\!2)$ bands. The in-band rotational spectrum is that of 
a rigid rotor with characteristic $L(L+1)$ splitting for all $K$-bands.
\begin{figure}[t]
\centering
\includegraphics[width=\linewidth,clip]{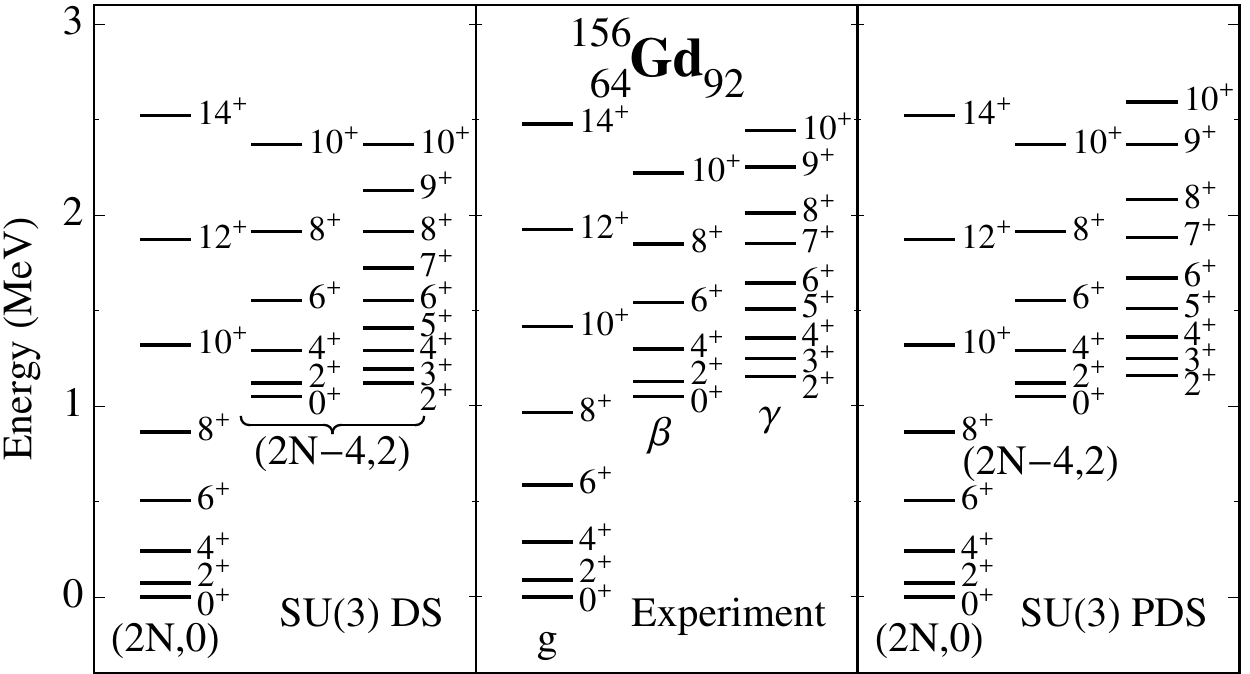}
\caption{
Observed spectrum of $^{156}$Gd~\cite{Reich03}
compared with the calculated spectra
of $\hat H_{\rm DS}$~(\ref{e_ds}) with SU(3) DS
and of $\hat H_{\rm PDS}$~(\ref{h_PDS-2}) with SU(3) PDS for $N=12$. 
The parameters are $\alpha_1=-7.6,\,\alpha_2=12.0,\,\alpha_3=0$ keV 
in both Hamiltonians and $\eta_1=-0.23,\, \eta_3=1.54$ keV in 
$\hat H_{\rm PDS}$. $(\lambda,\mu)$ are SU(3) labels. 
Adapted from~\cite{levramisa13}.}
\label{fig-1}
\end{figure}
\begin{figure}
\centering
\includegraphics[width=\linewidth,clip]{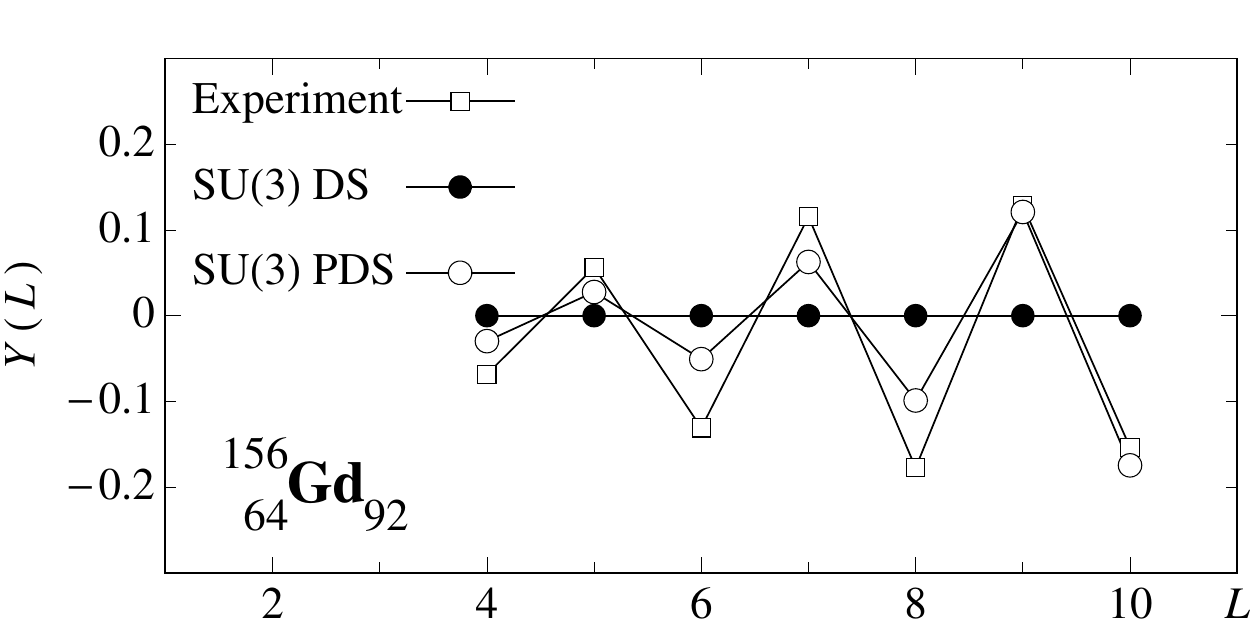}
\caption{
Observed~\cite{Reich03} and calculated (SU(3) DS and PDS) 
odd-even staggering of the $\gamma$ band in $^{156}$Gd. Adapted 
from~\cite{levramisa13}.}
\label{fig-2}
\end{figure}
\begin{table*}
\centering
\caption{\label{t_be2}
Observed~\cite{Reich03} and calculated $B$(E2) values in $^{156}$Gd.
For both the DS and PDS calculations, the E2 operator is
$e_{\rm b}[ s^\dag\tilde d +d^\dag s +\chi(d^\dag\tilde d)^{(2)}]$
with an effective boson charge $e_{\rm b}=0.166$~$eb$ and $\chi=-0.168$. 
Adapted from~\cite{levramisa13}.}
\begin{tabular}{llll|lllll}
\hline
Transition & Experiment & DS & PDS\hspace {0.5cm} & 
Transition & Experiment & DS & PDS\\
\hline
$  2^+_1\rightarrow0^+_1$& 0.933~{\sl 25}  &  0.933 & 0.933 &
$4^+_\beta\rightarrow2^+_1$& 0.0065~{\sl 35}  & 0.0067 &  0.0067 \\
$  4^+_1\rightarrow2^+_1$& 1.312~{\sl 25}  & 1.313 & 1.313 &
$4^+_\beta\rightarrow4^+_1$&~~~ ---  & 0.0067 & 0.0067 \\
$  6^+_1\rightarrow4^+_1$& 1.472~{\sl 40}  & 1.405 & 1.405 &
$4^+_\beta\rightarrow6^+_1$& 0.0105~{\sl 55} & 0.021 & 0.021 \\
$  8^+_1\rightarrow6^+_1$& 1.596~{\sl 85}  & 1.409 & 1.409 &
$2^+_\gamma\rightarrow0^+_1$& 0.0233~{\sl 8}  & 0.035 & 0.030 \\
$10^+_1\rightarrow8^+_1$& 1.566~{\sl 70}  & 1.364 & 1.364 &
$2^+_\gamma\rightarrow2^+_1$& 0.0361~{\sl 12}  & 0.056 & 0.048 \\
$2^+_\beta\rightarrow0^+_\beta$& 0.26~{\sl 11}  & 0.679 & 0.679 &
$2^+_\gamma\rightarrow4^+_1$& 0.0038~{\sl 2}  & 0.0037 & 0.0031 \\
$4^+_\beta\rightarrow2^+_\beta$& 1.40~{\sl 75}  & 0.951 &  0.951 &
$3^+_\gamma\rightarrow2^+_1$& 0.0364~{\sl 70}  & 0.062 & 0.053 \\
$0^+_\beta\rightarrow2^+_1$& 0.04~{\sl 2}  & 0.034 & 0.034 &
$3^+_\gamma\rightarrow4^+_1$& 0.0254~{\sl 50}  & 0.032 & 0.028 \\
$2^+_\beta\rightarrow0^+_1$& 0.0031~{\sl 3}  & 0.0055 & 0.0055 &
$4^+_\gamma\rightarrow2^+_1$& 0.0090~{\sl 25}  & 0.017 & 0.015 \\
$2^+_\beta\rightarrow2^+_1$& 0.0165~{\sl 15}  & 0.0084 & 0.0084 &
$4^+_\gamma\rightarrow4^+_1$& 0.050~{\sl 15}  & 0.067 & 0.057 \\
$2^+_\beta\rightarrow4^+_1$& 0.0204~{\sl 20}  & 0.020 & 0.020 &
$4^+_\gamma\rightarrow6^+_1$&~~~ ---  & 0.0089 & 0.0076 \\
                          &                  &       &       &
$4^+_\gamma\rightarrow2^+_\beta$& 0.0214~{\sl 80}  & 0.0033 & 0.0096\\
\hline
\end{tabular}
\end{table*}

A comparison with the experimental 
spectrum and E2 rates of $^{156}$Gd 
is shown in Figs.~\ref{fig-1}-\ref{fig-2} and Table~\ref{t_be2}. 
It displays 
a good description for properties of 
states in the ground and $\beta$ bands.
However, the resulting fit to energies of the $\gamma$-band 
is quite poor. The latter
display an odd-even staggering with pronounced deviations 
from a rigid-rotor pattern, indicative of a triaxial behavior. 
This effect can be visualized by plotting the quantity~\cite{Casten91}
\ba
Y(L)=
\frac{2L-1}{L}\times
\frac{E(L)-E(L-1)}{E(L)-E(L-2)}-1 ~,
\label{e_stag}
\ea
where $E(L)$ is the excitation energy
of a $\gamma$-band level with angular momentum $L$.
For a rotor this quantity is flat, $Y(L)=0$,
as illustrated in Fig.~\ref{fig-2} with the SU(3) DS calculation, 
in marked disagreement with the empirical data. 
A proper treatment of such signature splitting 
necessitates the inclusion of at least cubic terms in the
IBM Hamiltonian. 
In the IBM there are 17 possible three-body interactions.
One is thus confronted with the need to select suitable higher-order terms
that can break the DS in the $\gamma$-band but preserve it in the 
ground and $\beta$ bands.
These are precisely the defining 
properties of a partial dynamical symmetry (PDS)~\cite{Leviatan11}. 
The essential idea 
is to relax the stringent conditions of {\em complete} solvability, 
so that only part of the eigenspectrum retains all the DS quantum numbers.  
Various types of bosonic and fermionic PDS are known to be relevant to 
nuclear spectroscopy~\cite{Leviatan11,Leviatan96,levsin99,casten14,
levisa02,kremer14,escher00,isa14,Ramos09,levramisa13}. 
In the present contribution we demonstrate the relevance of PDS to 
the odd-even staggering in the $\gamma$-band of $^{156}$Gd~\cite{levramisa13}.

The method to construct Hamiltonians with PDS 
is based on an expansion in terms of tensors 
which annihilate prescribed set of states~\cite{Alhassid92,Ramos09}
\ba
\hat{H}_{\rm PDS} = 
\sum_{\alpha,\beta} 
u_{\alpha\beta}\, \hat{B}^{\dag}_{\alpha}\hat{B}_{\beta} ~. 
\label{h_PDS}
\ea
In the present study, the tensors involve
$n$-boson 
creation and annihilation 
operators $\hat{B}^{\dag}_{\alpha}\equiv 
\hat{B}^{\dagger}_{[n](\lambda,\mu)K;\ell m}$ 
and $\hat{B}_{\alpha}$, with definite character under the SU(3) 
chain~(\ref{e_clas}), 
which satisfy
\ba
\hat{B}_{\alpha}\vert [N](2N,0),K=0,L\rangle = 0 ~.
\label{Balpha}
\ea
Interactions as in Eq.~(\ref{h_PDS})
can be added to the Hamiltonian~(\ref{e_ds}) 
without destroying solvability of part of its spectrum. 
For $n=2$, we find two such SU(3) tensors
\bsub
\ba
\hat{B}^{\dagger}_{[2](0,2)0;00} &\propto& 
P^{\dagger}_{0} = d^{\dagger}\cdot d^{\dagger} - 2(s^{\dagger})^2,
\label{P0}\\
\hat{B}^{\dagger}_{[2](0,2)0;2m} &\propto& 
P^{\dagger}_{2m} = 2d^{\dagger}_{m}s^{\dagger} + 
\sqrt{7}\, (d^{\dagger}\,d^{\dagger})^{(2)}_{m}, 
\label{P2}
\ea
\label{PL}
\esub
with $(\lambda,\mu)=(0,2)$ and angular momentum $\ell=0,2$. 
For $n=3$, we find six SU(3) tensors 
\bsub
\ba
\hat B^\dag_{[3](2,2)0;00} 
&\propto&
W^{\dag}_{0} = 5P^{\dag}_{0}s^{\dag}- P^{\dag}_{2}\cdot d^{\dag} ~,\\
\hat B^\dag_{[3](2,2)2;2m} 
&\propto& 
W^{\dag}_{2m} = P^{\dag}_{0}d^{\dag}_{m} + 2P^{\dag}_{2m}s^{\dag} ~,
\label{W2}\\
\hat B^\dag_{[3](2,2)0;2m} 
&\propto& 
V^{\dag}_{2m} = 6P^{\dag}_{0}d^{\dag}_{m} - P^{\dag}_{2m}s^{\dag} ~,\\
\hat B^\dag_{[3](2,2)2;\ell m} 
&\propto& 
W^{\dag}_{\ell m} = (P^{\dag}_{2}d^{\dag})^{(\ell)}_{m}\;\;\; 
\ell=3,4 ~, 
\label{WV}\\
\hat B^\dag_{[3](0,0)0;00} 
&\propto& 
\Lambda^{\dag} = P^{\dag}_{0}s^{\dag} + P^{\dag}_{2}\cdot d^{\dag} ~,
\label{Lam}
\label{Lam0}
\ea
\label{PLB3}
\esub
with $(\lambda,\mu)=(2,2)$, $\ell=0,2,2,3,4$ and 
$(\lambda,\mu)=(0,0)$, $\ell=0$, all satisfying Eq.~(\ref{Balpha}) 

The aforementioned Casimir operators of SU(3) and the integrity basis 
operator $\hat{\Omega} = -4\sqrt{3}\, \hat{Q}\cdot 
(\hat{L}\times \hat{L})^{(2)}$, which conserve SU(3), 
are contained in the expression~(\ref{h_PDS}).
In general, however, $\hat{H}_{\rm PDS}$~(\ref{h_PDS}) 
breaks SU(3) yet, by construction, 
for {\em any} choice of parameters 
the ground-band members $|[N](2N,0)K\!=\!0,L\rangle$ are solvable 
with good SU(3) symmetry. 
For specific choices, additional solvable states are obtained, as we 
outline below.

Of particular interest is a class of Hamiltonians with SU(3) PDS 
which has solvable ground and $\beta$ bands 
with good SU(3) symmetry and energies $E_{\rm DS}$ (\ref{ene_ds}). 
This follows from the structure of the relevant Hamiltonian,
\ba
\hat H_{\rm PDS}=
\hat H_{\rm DS}+
\eta_2W_2^\dag\!\cdot\!\tilde W_2
+ \eta_3W_3^\dag\!\cdot\!\tilde W_3 ~,
\label{h_PDS-2}
\ea
and the fact that $W_{2m}$ (\ref{W2}) and $W_{3m}$ (\ref{WV}) satisfy 
Eq.~(\ref{Balpha}) and
\ba
W_{\ell m}\vert [N](2N-4,2),K=0,L\rangle = 0 \;\;\; \ell=2,3 ~.
\ea
The eigenstates of these solvable bands,
$|[N](2N,0)K\!=\!0,L\rangle$ and $|[N](2N-4,2,0)K\!=\!0,L\rangle$, 
are the same for the DS and PDS Hamiltonians.
In contrast, other bands including the $\gamma$ band 
are mixed with respect to SU(3).
As shown in Fig.~2, 
the empirical odd-even staggering in the $\gamma$-band 
in $^{156}$Gd is well reproduced by the PDS Hamiltonian (\ref{h_PDS-2}). 
The calculated staggering increases with $L$
which agrees with the experiment up to $L=10$.
For the PDS calculation, the wave functions of the states in 
the $\gamma$ band involve $15\%$ SU(3) admixtures into the dominant 
$(2N-4,2)$ component. 
These admixtures reflect 
the coupling of the $\gamma$ band with higher-lying excited bands.
Other approaches advocating the coupling of the $\gamma$ band to the 
$\beta$ band~\cite{Bonatsos88} or to the ground band~\cite{Minkov00}
fail to describe the odd-even staggering in $^{156}$Gd. 
Higher bands exhibit larger SU(3) mixing 
and their wave functions are spread over 
many SU(3) irreps, as shown for the $K=0_3$ band in Fig.~3. 
This complex SU(3) decomposition is 
in marked contrast to the SU(3)-purity of the ground ($K=0_1$) and 
$\beta$ ($K=0_2$) bands. 
\begin{figure}
\centering
\rotatebox{270}{\includegraphics[height=6.5cm]{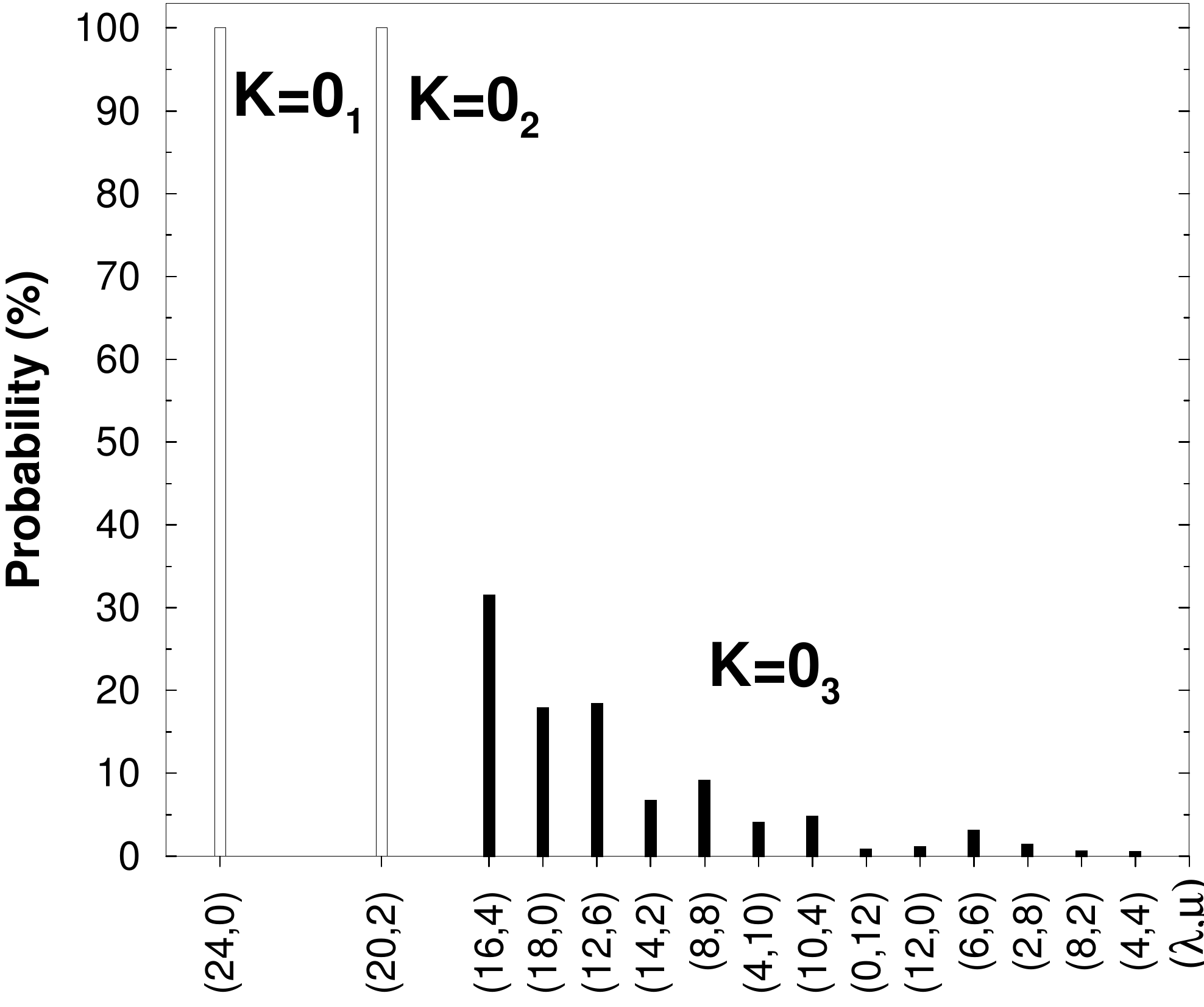}}
\caption{SU(3) decomposition of wave functions of $L\!=\!0$ states 
in the $K=0_1,\,0_2,\,0_3$ bands for the PDS calculation.
Adapted from \cite{levramisa13}.}
\label{fig-3}
\end{figure}

It should be emphasized that the PDS results for the $\gamma$ band 
are obtained without altering the good agreement for the ground and 
$\beta$ bands, already achieved with the SU(3) DS calculation. 
This is further illustrated with the E2 transitions in $^{156}$Gd.
The observed $B$(E2) values between ground, $\beta$, and $\gamma$ bands 
are shown in Table~\ref{t_be2}
and compared to the results of the SU(3) DS and PDS calculations.
The E2 transitions between ground and $\beta$ bands
can be calculated analytically, and remain valid in SU(3) PDS.
Transitions involving $\gamma$-band members
are different in SU(3) DS and PDS, 
and are computed numerically for the latter. 
It is seen from Table~\ref{t_be2} that the mixing of the $\gamma$ band 
with higher-lying excited bands improves the agreement with the data 
in most cases.

Another class of Hamiltonians with SU(3) PDS exists
which has solvable members of $\gamma^{k}(K\!=\!2k)$ bands 
$|[N](2N-4k,2k)K\!=\!2k,L\rangle$, with energies 
$E_{\rm DS}$ (\ref{ene_ds}).
This follows from the structure of the relevant Hamiltonian,
\ba
\hat H'_{\rm PDS}
&=& \hat{H}_{DS} 
+ t_{0}P^{\dag}_{0}P_{0} 
+ u_{0}P^{\dag}_{0}s^{\dag}sP_{0}\nonumber\\
&&
+ v_{0}\left ( \Lambda^{\dag}sP_{0} + 
 P^{\dag}_{0}s^{\dag}\Lambda\right ) ~,
\label{hPDS-1}
\ea
and the fact that $P_0$ (\ref{P0}) and $\Lambda$ (\ref{Lam0}) satisfy 
Eq.~(\ref{Balpha}) and
\bsub
\ba
&&P_{0}\,\vert [N](2N-4k,2k)K=2k, L\rangle = 0 ~,\\
&&\Lambda\,\vert [N](2N-4k,2k)K, L\rangle = 0 ~.
\ea
\esub
Two-body Hamiltonians of this class have been 
shown to play a role in diverse phenomena, including 
spectroscopy of rare-earth nuclei~\cite{Leviatan96,levsin99,casten14}, 
quantum phase transitions~\cite{lev07,maclev14} and 
mixed regular and chaotic dynamics~\cite{maclev14,WAL93}.
The two classes of SU(3)-PDS Hamiltonians demonstrate 
the increase in flexibility obtained by generalizing the concept of DS to PDS. 
In fact, in the IBM more than half of all possible interactions have an 
SU(3) PDS~\cite{levramisa13}. 

In summary, we have presented several classes of 
IBM Hamiltonians with SU(3) PDS, and obtained an improved 
description of signature splitting in the $\gamma$ band of $^{156}$Gd. 
The analysis serves to highlight the merits gained by
using the notion of PDS as a tool for selecting 
higher-order terms in systems where 
a prescribed symmetry is not obeyed uniformly. 
On one hand, the PDS approach allows more flexibility 
by relaxing the constraints of an exact DS. 
On the other hand, the PDS picks particular symmetry-breaking terms 
without destroying results previously 
obtained with a DS for a segment of the spectrum. 
The non-solvable states can experience strong symmetry-breaking. 
These virtues can be exploited in studying the role of higher-order terms 
in collective Hamiltonians and in attempts to extend 
beyond-mean-field methods to heavy nuclei.

This work was done in collaboration 
with J.E.~Garc\'\i a-Ramos (Huelva) and P. Van Isacker (GANIL) 
and is supported by the Israel Science Foundation.

\end{document}